\documentclass[reprint,aps,superscriptaddress,onecolumn]{revtex4-2}

\usepackage[T1]{fontenc}
\usepackage[utf8]{luainputenc}
\usepackage[english]{babel}
\usepackage{float}
\usepackage{bm}
\usepackage{amsmath}
\usepackage{amsthm}
\usepackage{amssymb}
\usepackage{graphicx}
\usepackage{algorithm,algorithmic}
\usepackage[unicode=true]
{hyperref}

\theoremstyle{definition}
\newtheorem{defn}{\protect\definitionname}
\providecommand{\definitionname}{Definition}

\begin{document}


\title{Vortex boundaries as barriers to diffusive vorticity transport in two-dimensional flows}
\author{Stergios Katsanoulis}
\affiliation{Institute for Mechanical Systems, ETH Zurich, Leonhardstrasse 21, 8092 Zurich, Switzerland}
\author{Mohammad Farazmand}
\affiliation{Department of Mathematics, North Carolina State University, 2311
	Stinson Dr., Raleigh, NC 27695, USA}
\author{Mattia Serra}
\affiliation{School of Engineering and Applied Sciences, Harvard University,
	29 Oxford Street, Cambridge, MA 02138, USA}
\author{George Haller}
\email{georgehaller@ethz.ch}
\affiliation{Institute for Mechanical Systems, ETH Zurich, Leonhardstrasse 21, 8092 Zurich, Switzerland}



\date{\today}

\begin{abstract}
We put forward the idea of defining vortex boundaries in planar flows
as closed material barriers to the diffusive transport of vorticity. Such diffusive vortex boundaries minimize the leakage of vorticity from the fluid mass they enclose
when compared to other nearby material curves. Building on recent
results on passive diffusion barriers, we develop an algorithm for
the automated identification of such structures from general, two-dimensional unsteady flow data. As examples, we identify vortex boundaries as vorticity diffusion barriers in two flows: an explicitly known laminar flow and a numerically generated turbulent Navier--Stokes flow.
\end{abstract}


\maketitle

\section{Introduction}
Vortices in turbulent flows are omnipresent yet difficult
to define unambiguously. As argued by \cite{haller16}, however, two
common expectations for vortices have emerged in the literature: Material
invariance and high levels of vorticity. 

Regarding material invariance, Lugt \cite{lugt1979dilemma}
expects a vortex to be formed by material particles rotating around
a common center, while McWilliams \cite{mcwilliams_1984} requires
a vortex to ``persist under passive advection by the large-scale
flow''. Chong et al. \cite{chong1990general} view vortices as sets
of instantaneously spiraling particle motions. Provenzale \cite{provenzale99}
emphasizes small material dispersion within vortex cores \cite{cucitore1999effectiveness}.
Chakraborty et al. \cite{chakraborty2005relationships} argue that
both swirling motion and small particle dispersion are important features
of a vortex core. Haller \cite{haller2005objective} views vortices
as sets of non-hyperbolic trajectories and Chelton et al. \cite{chelton2011global}
postulate that nonlinear eddies trap and carry fluid in their interior.
In a similar setting, Mason et al. \cite{mason2014new} seek vortices
that are ``efficient carriers of mass and its physical, chemical,
and biological properties''.

Regarding vorticity in a vortex, McWilliams \cite{mcwilliams_1984}
and Hussein \cite{hussain1986coherent} expect high vorticity in vortices
compared with the background flow. In contrast, Okubo \cite{okubo1970},
Hunt et al. \cite{hunt1988eddies}, Weiss \cite{weiss1991}, Hua \&
Klein \cite{hua1998exact} and Hua et al. \cite{hua1998lagrangian}
require vorticity to dominate strain inside a vortex. Others compare
vorticity to strain in the rate-of-strain eigenbasis \cite{tabor1994stretching}
\cite{lapeyre1999does} \cite{lapeyre2001comment}. Further variants
of these ideas have been developed in the scientific visualization
community, as reviewed in \cite{gunther2018state}.

Formulating these guiding principles into a simple vortex
definition has been a major challenge. At a conceptual level, the
required material nature of the vortex necessitates an approach that
truly targets material behavior. A litmus test for self-consistent
material description is independence of the observer (or \emph{objectivity}),
which has long been enforced in continuum mechanics \cite{truesdell2004non}
\cite{gurtin1982introduction} \cite{gurtin2010mechanics} for any
theory purporting to describe material response. Objectivity was also
identified as a basic requirement for flow feature detection in fluid
mechanics already in the 1970's \cite{drouot1976definition} \cite{drouot1976approximation}
\cite{lugt1979dilemma}, yet objective Lagrangian criteria for material
vortex boundaries in two-dimensional flows have only appeared in recent
years \cite{haller2013coherent} \cite{farazmand2016polar} \cite{haller16}.
Of these approaches, only \cite{haller16} involves the vorticity
as a kinematic measure of rotational coherence. Seeking material regions
from which vorticity transport is minimal, however, requires the involvement
of the Navier--Stokes equations, an element that has been missing
in the purely kinematic vortex criteria survived above. 

More generally, finding theoretically optimal barriers to
the transport of diffusive quantities in fluid flows has been an elusive
problem (see, e.g., \cite{weiss2007transport}). As a recent advance
in this area, \cite{haller18-1} formulated and solved a precise variational
problem for material surfaces that inhibit the transport of weakly
diffusive, passive scalar fields more than neighboring surfaces do
in an incompressible flow. These results have subsequently been extended
to compressible flows and to scalar fields with a known (and hence
constrained) initial concentration \cite{haller18-2}.

Once a two-dimensional incompressible velocity field is
known, its associated vorticity transport equation becomes a scalar
advection-diffusion equation for the scalar vorticity. The initial
condition of the equation, however, is constrained to be the plane-normal
component of the curl of the velocity field at the initial time. Therefore,
the general constrained transport barrier results of \cite{haller18-2}
apply to vorticity transport in incompressible, planar Navier--Stokes
flows. We exploit this fact here and invoke the results of \cite{haller18-2}
to define and locate closed material curves that inhibit the leakage
of vorticity from their interior most effectively.

We construct diffusive vortex boundaries as outermost
periodic orbits of an explicit ordinary differential equation family
arising from the exact solution to the minimal vorticity leakage problem.
This automated algorithm is now publicly available under \href{https://github.com/katsanoulis/BarrierTool}{https://github.com/katsanoulis/BarrierTool}
in the MATLAB \textit{\emph{package entitled}}\textit{ BarrierTool}.
We illustrate this algorithm first on an explicitly known solution
of the planar Navier-Stokes equations, then on a two-dimensional decaying
turbulence simulation.

\section{Constrained material barriers to vorticity transport }
As mentioned in the Introduction, Haller et al.~\cite{haller18-2} have derived the criteria
for locating material barriers to diffusive transport in compressible
flows. These results are applicable to arbitrary passive scalar fields in arbitrary
spatial dimensions and with arbitrary diffusion tensors that possibly
depend on space and time. Concentration sinks and sources, as well
as spontaneous concentration decay are also allowed. Here we recall
these results specifically formulated for the two-dimensional scalar
vorticity field $\omega(\mathbf{x},t)$ of an incompressible, two-dimensional
Navier--Stokes velocity field $\mathbf{v}(\mathbf{x},t)$ whose kinematic
viscosity is $\nu\geq0$. 

In this context, if $\mathbf{v}(\mathbf{x},t)$ is known,
then $\omega(\mathbf{x},t)$ satisfies the two-dimensional, linear
advection-diffusion equation
\begin{align}
\partial_{t}\omega+\mathbf{\bm{\nabla}}\omega\cdot\mathbf{v} & =\nu\Delta\omega,\label{eq:adv-diff}\\
\omega(\mathbf{x},t_{0}) & =\omega_{0}(\mathbf{x}),\nonumber 
\end{align}
where $\Delta$ is the Laplacian operator and $\bm{\mathbf{\nabla}}$
denotes the gradient operation with respect to the spatial variable
$\mathbf{x}\in U\subset\mathbb{R}^{2}$ on a compact domain $U$.
We denote the flow map generated by the trajectories \textbf{$\mathbf{x}(t;t_{0},\mathbf{x}_{0})$
}of the velocity $\mathbf{v}(\mathbf{x},t)$ by\textbf{ $\mathbf{F}_{t_{0}}^{t}\left(\mathbf{x}_{0}\right):=\mathbf{x}(t;t_{0},\mathbf{x}_{0})$}.
Consider an evolving material curve $\mathcal{M}(t)=\mathbf{F}_{t_{0}}^{t}\left(\mathcal{M}_{0}\right)$
with initial position $\mathcal{M}(t_{0})=\mathcal{M}_{0}$. Let $s\in\left[\alpha,\beta\right]$
denote a parametrization of $\mathcal{M}_{0}$ and let $\mathbf{n}_{0}(s)$
denote a smooth unit normal vector field along $\mathcal{M}_{0}$,
and let $\Sigma_{t_{0}}^{t_{1}}(\mathcal{M}_{0})$ denote the total
normed transport of $\omega$ through the material curve $\mathcal{M}(t)$
over the time interval $[t_0,t_1]$. By normed transport we mean the time integral
of the normed instantaneous flux, which therefore sums up all the
leakage of $\omega$ through a curve without cancellations. The quantity
$\Sigma_{t_{0}}^{t_{1}}(\mathcal{M}_{0})$ is ideal for assessing
the permeability of a surface for transport, whereas the unnormed
(signed) vorticity transport may be small due to cancellations even
for a highly permeable material curve. Note that both the normed and
the unnormed vorticity transport are purely diffusive (i.e., vanish
for $\nu=0$), given that $\mathcal{M}(t)$ is a material surface
and hence blocks all advective transport of a passive scalar field.

We are interested in finding initial curves $\mathcal{M}_{0}$
that extremize the normed and normalized vorticity transport
functional 
\[
\tilde{\Sigma}_{t_{0}}^{t_{1}}(\mathcal{M}_{0}):=\frac{\Sigma_{t_{0}}^{t_{1}}(\mathcal{M}_{0})}{\nu\left(t_{1}-t_{0}\right)\int_{\mathcal{M}_{0}}ds},
\]
where we have normalized the normed transport by the diffusivity,
the length of the time interval and the length of the material curve
$\mathcal{M}_{0}$. As shown by \cite{haller18-2}, $\tilde{\Sigma}_{t_{0}}^{t_{1}}(\mathcal{M}_{0})$
can be rewritten as
\[
\tilde{\Sigma}_{t_{0}}^{t_{1}}(\mathcal{M}_{0})=\frac{\int_{\mathcal{M}_{0}}\left|\left\langle \mathbf{\bar{q}}_{t_{0}}^{t_{1}}(\mathbf{x}_{0}(s)),\mathbf{n}_{0}\left(s\right)\right\rangle \right|\,ds}{\int_{\mathcal{M}_{0}}ds},
\]
where $\mathbf x_0(s)$ denotes the parameterization of $\mathcal M_0$ ($s\in[\alpha,\beta]$)
and the \emph{transport vector field} $\mathbf{\bar{q}}_{t_{0}}^{t_{1}}(\mathbf{x}_{0})$
is given by 
\begin{align}
\mathbf{\bar{q}}_{t_{0}}^{t_{1}}(\mathbf{x}_{0}) & =\frac{1}{t_{1}-t_{0}}\int_{t_{0}}^{t_{1}}\left[\mathbf{\bm{\nabla}}_{0}\mathbf{F}_{t_{0}}^{t}(\mathbf{x}_{0})\right]^{-1}\left[\mathbf{\bm{\nabla}}\omega\left(\mathbf{F}_{t_{0}}^{t}\left(\mathbf{x}_{0}\right),t\right)\right]dt,
\label{eq:q def equiv}
\end{align}
where $\bm\nabla_0$ denotes the derivative with respect to $\mathbf x_0$.

Material curves, $\mathcal{M}(t)$, that extremize $\tilde{\Sigma}_{t_{0}}^{t_{1}}$
have initial positions for which the variational derivative of $\tilde{\Sigma}_{t_{0}}^{t_{1}}$
vanishes,
\begin{equation}
\delta\mathcal{\tilde{E}}(\mathcal{M}_{0}^{*})=0.\label{eq:variproblem_known_IC-1}
\end{equation}
Haller et al. \cite{haller18-2} have obtained that the most observable class of
solutions of this variational problem, \emph{uniform vorticity barriers},
satisfy the conservation law 
\begin{equation}
\left|\left\langle \mathbf{\bar{q}}_{t_{0}}^{t_{1}}(\mathbf{x}_{0}(s)),\mathbf{n}_{0}\left(s)\right)\right\rangle \right|=\mathcal{T}_{0},\qquad0\leq\mathcal{T}_{0}\leq\max_{\mathbf{x}_{0}\in U}\left|\mathbf{\bar{q}}_{t_{0}}^{t_{1}}(\mathbf{x}_{0})\right|,
\label{eq:uniform barriers}
\end{equation}
for some constant $\mathcal{T}_{0}$, which measures the pointwise
constant, uniform transport density along such barriers. This conservation
law gives an implicit differential equation for curve families $\mathbf{x}_{0}(s)$
that span initial positions of uniform material barriers to the diffusive
transport of $\omega$. Haller et al.~\cite{haller18-2} also show that an explicit
differential equation family equivalent to the implicit one in (\ref{eq:uniform barriers})
is given by 
\begin{equation}
\mathbf{x}_{0}^{\prime}=\left(\mathcal{T}_{0}\mathbf{\bm{\Omega}}\pm\sqrt{\left|\mathbf{\bar{q}}_{t_{0}}^{t_{1}}\left(\mathbf{x}_{0}\right)\right|^{2}-\mathcal{T}_{0}^{2}}\,\mathbf{I}\right)\int_{t_{0}}^{t_{1}}\left[\mathbf{\bm{\nabla}}_{0}\mathbf{F}_{t_{0}}^{t}(\mathbf{x}_{0})\right]^{-1}\mathbf{\bm{\nabla}}\omega\left(\mathbf{F}_{t_{0}}^{t}\left(\mathbf{x}_{0}\right),t\right)dt,\qquad\mathbf{\bm{\Omega}}:=\left(\begin{array}{rc}
0 & 1\\
-1 & 0
\end{array}\right).\label{eq:barrier ODE}
\end{equation}
Finally, Ref. \cite{haller18-2} obtains a scalar diagnostic field, the
\emph{diffusion barrier strength} (DBS) field, that measures the local
strength of transport barriers. This barrier strength is equal to
the leading-order change in the local transport under small, localized
normal perturbations to a transport barrier. The DBS field can simply
be computed as

\begin{equation}
{\displaystyle DBS_{t_{0}}^{t_{1}}(\mathbf{x}_{0})=\left|\bar{\mathbf{q}}_{t_{0}}^{t_{1}}(\mathbf{x}_{0})\right|,}\label{eq:DBS field}
\end{equation}
with its ridges delineating the most influential vorticity
transport extremizers. Both the exact differential equation (\ref{eq:barrier ODE})
and the diagnostic field $DBS_{t_{0}}^{t_{1}}$ are objective, as
shown by \cite{haller18-2}.

\section{Diffusive vortex boundaries as closed material barriers to vorticity transport }
The general equation (\ref{eq:barrier ODE}) for planar vorticity
barriers enables us to give a precise mathematical definition and
a computational algorithm for diffusive vortex boundaries as most observable
material inhibitors of vorticity leakage from a closed fluid region.
\begin{defn}
	A \emph{diffusive vortex boundary} over a time interval $[t_{0},t_{1}]$ is
	a closed material curve $\mathcal{M}^{*}(t)$ whose initial position
	$\mathcal{M}_{0}^{*}=\mathcal{M}^{*}(t_{0})$ is the outermost member
	of a closed orbit family in the differential equation family (\ref{eq:barrier ODE}).
\end{defn}

\begin{figure}[H]
	\begin{centering}
		\includegraphics[width=0.4\textwidth]{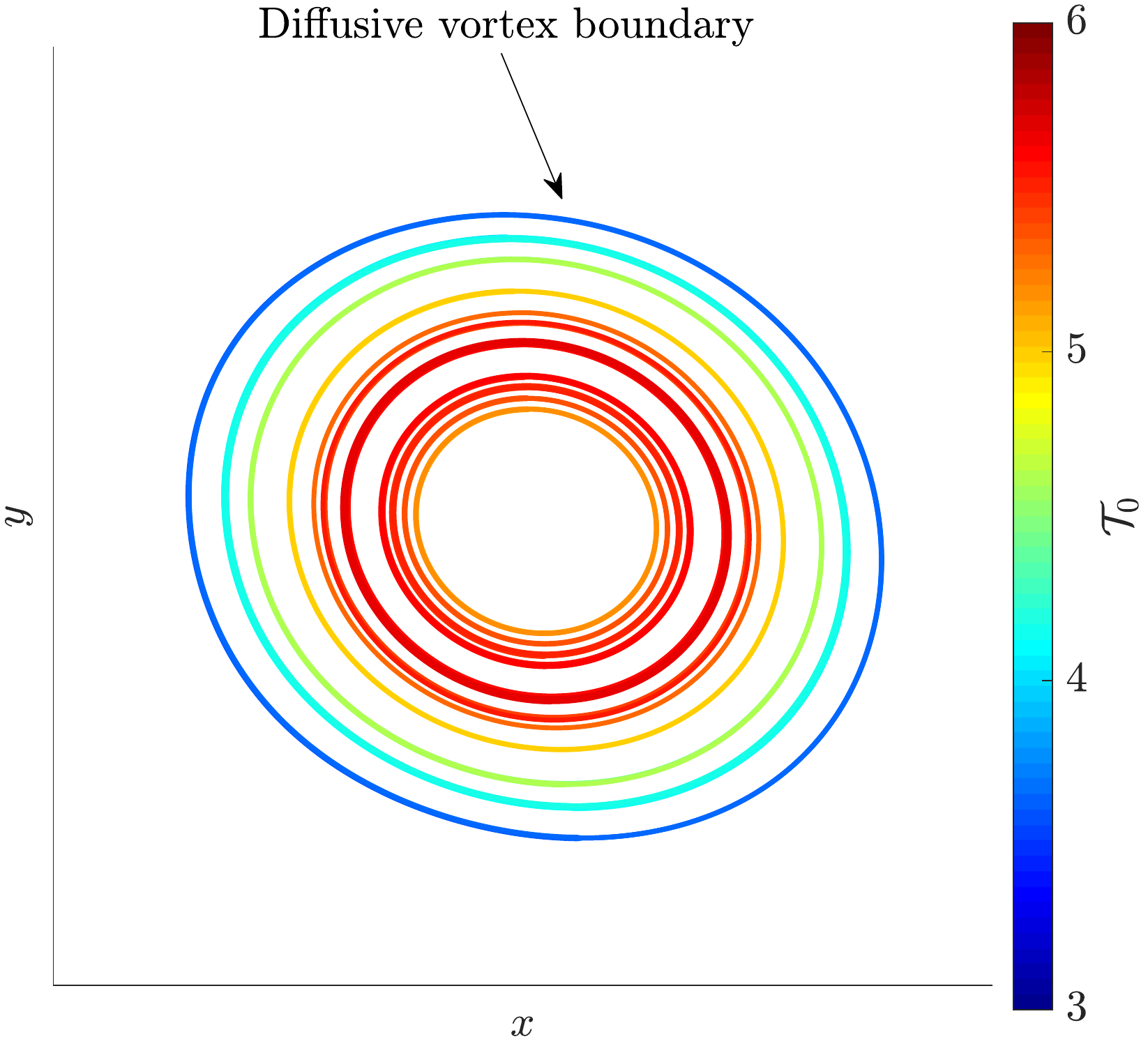}
		\par\end{centering}
	\caption{Family of limit cycles of equation \ref{eq:barrier ODE}. Each material curve has the same pointwise vorticity transport density $\mathcal{T}_{0}$. The outermost member of this family serves as the diffusive vortex boundary. \label{fig:vortexFamily}}
\end{figure}

Each member of a periodic orbit family in (\ref{eq:barrier ODE})
is technically a closed transport extremizer within its class. The
orbit family as a whole provides an internal stratification of a vortical
region into curves with the same uniform vorticity transport through
them. The outermost member of such a family is the practically observed
boundary of a region from which the leakage of vorticity is minimal, as shown in Fig. \ref{fig:vortexFamily}.
The material curve $\mathcal{M}^{*}(t)=\mathbf{F}_{t_{0}}^{t}\left(\mathcal{M}_{0}^{*}\right)$
in Definition 1 is fully determined by its initial position $\mathcal{M}_{0}^{*}$,
and hence the definition yields evolving material vortex boundaries
$\mathcal{M}^{*}(t)$ over the whole time interval $\left[t_{0},t_{1}\right]$.
The pointwise strength of such a diffusive vortex boundary can then be assessed
by computing $DBS_{t_{0}}^{t_{1}}(\mathbf{x}_{0})$ along its points. 

Given that $\mathbf{n}_{0}(s)=\mathbf{\Omega}\mathbf{x}_{0}^{\prime}(s)/\sqrt{\left\langle \mathbf{x}_{0}^{\prime}(s),\mathbf{x}_{0}^{\prime}(s)\right\rangle }$
is a smooth unit normal vector along any curve $\mathbf{x}_{0}(s)$,
parametrized curves satisfying the conservation law (\ref{eq:uniform barriers})
are also contained in the zero level set of the function family

\begin{equation}
{\displaystyle L\left(\mathbf{x}_{0},\mathbf{x}_{0}^{\prime};\mathcal{T}_{0}\right)=\sqrt{\left\langle \bar{\mathbf{q}}_{t_{0}}^{t_{1}}(\mathbf{x}_{0}),\mathbf{\Omega}\mathbf{x}_{0}'\right\rangle ^{2}}-\mathcal{T}_{0}\sqrt{\left\langle \mathbf{x}_{0}',\mathbf{x}_{0}'\right\rangle },}\label{eq:Lagrangian}
\end{equation}
which also turns out to be the Lagrangian associated with the variational
problem (\ref{eq:variproblem_known_IC-1}) (cf. Ref. \cite{haller18-2})
To locate closed zero-level curves of $L$, we adapt an idea originally developed
in \cite{serra17} for the automated computation of null-geodesics.
First, we observe that the conservation law (\ref{eq:uniform barriers}),
and accordingly the zero level set of $L$, is invariant under reparametrizations
of the curve $\mathbf{x}_{0}(s)$. This enables us to parametrize
the yet unknown $\mathcal{M}_{0}^{*}$ by arclength, i.e., for an
appropriate angle $\varphi(s)$, we can set 

\begin{equation}
\mathbf{x}{}_{0}^{\prime}(s)=\boldsymbol{e}_{\varphi}(s):=\left(\begin{array}{c}
\cos\varphi(s)\\
\sin\varphi(s)
\end{array}\right),\quad\implies\mathbf{x}_{0}^{\prime\prime}=\boldsymbol{e}_{\varphi}+\mathbf{\bm{\Omega}}^{T}\boldsymbol{e}_{\varphi}\varphi'.\label{eq:ephi def}
\end{equation}

Thus, by (\ref{eq:Lagrangian}), curves in the zero level
set of $L$ satisfy

\begin{equation}
\sqrt{\left\langle \bm{\Omega}\mathbf{\bar{q}}_{t_{0}}^{t_{1}}\left(\mathbf{x}_{0}(s)\right),\boldsymbol{e}_{\varphi}(s)\right\rangle ^{2}}-\mathcal{T}_{0}=0.\label{eq:zero_level_set}
\end{equation}

Differentiating this last identity with respect to the parameter
$s$ and using the expression for $\mathbf{x}_{0}^{\prime\prime}$
from (\ref{eq:ephi def}) gives

\begin{equation}
\left\langle \bm{\Omega}\boldsymbol{\nabla}_{\mathbf{x}_{0}}\left(\mathbf{\bar{q}}_{t_{0}}^{t_{1}}\left(\mathbf{x}_{0}\right)\right)\boldsymbol{e}_{\varphi},\boldsymbol{e}_{\varphi}\right\rangle +\left\langle \mathbf{\bar{q}}_{t_{0}}^{t_{1}},\boldsymbol{e}_{\varphi}\right\rangle \varphi'=0.\label{eq:phiprime}
\end{equation}

Therefore, the definition of $\boldsymbol{e}_{\varphi}$
in (\ref{eq:ephi def}) and eq. (\ref{eq:phiprime}) together yield
an explicit, three-dimensional system of differential equations
\begin{align}
\mathbf{x}{}_{0}^{\prime} & =\boldsymbol{e}_{\varphi},\nonumber \\
\varphi^{\prime} & =\frac{\left\langle \bm{\Omega}\boldsymbol{\nabla}_{\mathbf{x}_{0}}\left(\mathbf{\bar{q}}_{t_{0}}^{t_{1}}\left(\mathbf{x}_{0}\right)\right)\boldsymbol{e}_{\varphi},\mathbf{e_{\varphi}}\right\rangle }{\left\langle \mathbf{\bar{q}}_{t_{0}}^{t_{1}}\left(\mathbf{x}_{0}\right),\boldsymbol{e}_{\varphi}\right\rangle },\label{eq:3D eq. for barriers}
\end{align}
defined on the set ${\displaystyle V=\left\{ \left(\mathbf{x}_{0},\varphi\right)\in U\times S^{1}:\left\langle \mathbf{\bar{q}}_{t_{0}}^{t_{1}}\left(\mathbf{x}_{0}\right),\boldsymbol{e}_{\varphi}\right\rangle \neq0\right\} }$.
Initial positions of closed material barriers to vorticity transport
are closed projections of trajectories of (\ref{eq:3D eq. for barriers})
to the plane of the $\mathbf{x}_{0}$ variable. 

\section{Numerical algorithm for diffusive vortex-boundary detection\label{sec:Numerical-algorithm}}
Geometrically, the original variational problem (\ref{eq:variproblem_known_IC-1})
leads to a four-dimensional system of ODEs in the space of the $(\mathbf{x}_{0},\mathbf{x}_{0}^{\prime})$
variables. The conservation law (\ref{eq:uniform barriers}) enables
us to reduce this four-dimensional ODE to the three-dimensional system
\eqref{eq:3D eq. for barriers}. Note that if a trajectory of (\ref{eq:3D eq. for barriers})
projects to a closed curve $\mathbf{x}_{0}(s)$, then for any angle
$\varphi_{0}\in \mathbb{S}^{1}$, there will be at least two points along the
curve where $\mathbf{x}{}_{0}^{\prime}=\boldsymbol{e}_{\varphi_{0}}=\left(\cos\varphi_{0},\cos\varphi_{0}\right)^{T}$.
Therefore, for any choice of $\varphi_{0}$, the set

\begin{equation}
{\displaystyle \mathcal{C}_{\mathcal{T}_{0}}=\left\{ \left(\mathbf{x}_{0},\varphi\right)\in U\times\left\{ \varphi_{0}\right\} :\left|\left\langle \bar{\mathbf{q}}_{t_{0}}^{t_{1}}(\mathbf{x}_{0}),\boldsymbol{e}_{\varphi_{0}}\right\rangle \right|=\mathcal{T}_{0}\right\} ,}\label{eq:C_T_0}
\end{equation}
is a set of curves that all periodic
orbits of (\ref{eq:3D eq. for barriers}) must cross. We can, therefore,
use the set $\mathcal{C}_{\mathcal{T}_{0}}$ as a Poincaré section
within each two-dimensional level set $\left|\left\langle \bar{\mathbf{q}}_{t_{0}}^{t_{1}}(\mathbf{x}_{0}),\boldsymbol{e}_{\varphi}\right\rangle \right|=\mathcal{T}_{0}$
to locate periodic orbits of (\ref{eq:3D eq. for barriers}). The
trivial choice for the angle $\varphi_{0}$ is $\varphi_{0}=0,$
in which case $\left\langle \bar{\mathbf{q}}_{t_{0}}^{t_{1}}(\mathbf{x}_{0}),\boldsymbol{e}_{\varphi_{0}}\right\rangle $
is simply the first component $\left[\bar{\mathbf{q}}_{t_{0}}^{t_{1}}(\mathbf{x}_{0})\right]_{1}$
of the vector $\bar{\mathbf{q}}_{t_{0}}^{t_{1}}(\mathbf{x}_{0})$
in the coordinate system selected for the analysis. These considerations lead to the numerical Algorithm \ref{alg:closedVortexBoundary} for
locating diffusive vortex boundaries in a two-dimensional Navier--Stokes flow.
\begin{figure}
\begin{algorithm}[H]
	\caption{Computing diffusive vortex boundaries}
	\label{alg:closedVortexBoundary}
	\begin{enumerate}
		\item Input the 2D velocity field $\mathbf{v}(\mathbf{x},t)$ defined over
		the spatial domain $U$ and time interval $[t_{0},t_{1}]$.
		\item Compute trajectories $\mathbf{x}(t;t_{0},\mathbf{x}_{0})$ of $\mathbf{v}(\mathbf{x},t)$
		over $[t_{0},t_{1}]$, starting from an initial grid $\mathcal{G}_{0}\subset U$.
		\item Calculate the deformation gradient $\mathbf{\bm{\nabla}}_{0}\mathbf{F}_{t_{0}}^{t}(\mathbf{x}_{0})$
		for $\mathbf{x}_{0}\in\mathcal{G}_{0}$ from finite differencing.
		Also, compute the vorticity gradient $\mathbf{\bm{\nabla}}\omega(\mathbf{x},t)$
		from finite differencing along each trajectory $\mathbf{x}(t;t_{0},\mathbf{x}_{0})$.
		Subsequently, compute the transport vector field $\mathbf{\bar{q}}_{t_{0}}^{t_{1}}\left(\mathbf{x}_{0}\right)$
		and its gradient from finite differencing to obtain the right-hand
		side of (\ref{eq:3D eq. for barriers}) over the grid $\mathcal{G}_{0}.$
		\item Fix a unit vector $\boldsymbol{e}_{\varphi}^{0}$ and set up a loop
		over values of the transport density constant $\mathcal{T}_{0}$ falling
		in the interval given in (\ref{eq:uniform barriers}).
		\item For each $\mathcal{T}_{0}$ value, calculate the initial condition
		set $\mathcal{C}_{\mathcal{T}_{0}}$ defined in (\ref{eq:C_T_0})
		by computing the level set $\left|\left\langle \bar{\mathbf{q}}_{t_{0}}^{t_{1}}(\mathbf{x}_{0}),\boldsymbol{e}_{\varphi}^{0}\right\rangle \right|=\mathcal{T}_{0}$.
		Launch trajectories of the ODE (\ref{eq:3D eq. for barriers}) from
		$\mathcal{C}_{\mathcal{T}_{0}}$. For off-the-grid points in $\mathcal{G}_{0}$,
		use bilinear interpolation to evaluate the right-hand side of eq.
		(\ref{eq:3D eq. for barriers}) for trajectory integration.
		\item Once the full loop of $\mathcal{T}_{0}$ is complete, identify diffusive vortex boundaries as the outermost members of the closed trajectory families
		obtained from the above procedure.
	\end{enumerate}
\end{algorithm}
\end{figure}

In our examples, we will use a MATLAB implementation of
the above algorithm, which is publicly available under \href{https://github.com/katsanoulis/BarrierTool}{https://github.com/katsanoulis/BarrierTool}.
This MATLAB package\textit{BarrierTool}, is in fact a more general
software tool that allows for the computation of elliptic Lagrangian
coherent structures \cite{haller2013coherent}, closed unconstrained
diffusion barriers \cite{haller18-1} and objective Eulerian coherent
structures \cite{serra2016objective}. 

We note that a more technical, alternative approach to solving
such variational problems involves a reduction of the original variational
problem to a two-dimensional direction field family \cite{haller18-2}.
Locating closed curves of this direction field family involves the
identification and analysis of direction field singularities. The
associated challenges are described in \cite{serra17} \cite{karrasch2015automated}. Recent progress on addressing some of these challenges is reported in \cite{karrasch2019fast}.

\section{Examples}
\subsection{Periodic array of recirculation cells}
A spatially periodic, steady solution of the 2D Euler equations
is given by \cite{majda2002vorticity}
\begin{equation}
\mathbf{v}_{E}(\mathbf{x})=2\left(\begin{array}{c}
-2\sin(2\pi x+4\pi y)-\sin(4\pi x+2\pi y)+\sin(4\pi x-2\pi y)+2\sin(2\pi x-4\pi y)\\
\sin(2\pi x+4\pi y)+2\sin(4\pi x+2\pi y)+2\sin(4\pi x-2\pi y)+\sin(2\pi x-4\pi y)
\end{array}\right).\label{eq:euler array-1}
\end{equation}
This Euler solution gives rise to the following spatially periodic, unsteady
solution of the 2D Navier--Stokes equations
\begin{equation}
\mathbf{v}(\mathbf{x},t)=e^{-20\pi^{2}\nu t}\mathbf{v}_{E}(\mathbf{x}),\label{eq:NS array}
\end{equation}
whose vorticity field 
\[
\omega(\mathbf{x},t)=20\pi e^{-20\pi^{2}\nu t}\left[\cos(2\pi x+4\pi y)+\cos(4\pi x+2\pi y)+\cos(4\pi x-2\pi y)+\cos(2\pi x-4\pi y)\right]
\]
satisfies the advection-diffusion equation (\ref{eq:adv-diff}) with
viscosity $\nu$.

\begin{figure}
	\begin{centering}
		\includegraphics[width=0.6\textwidth]{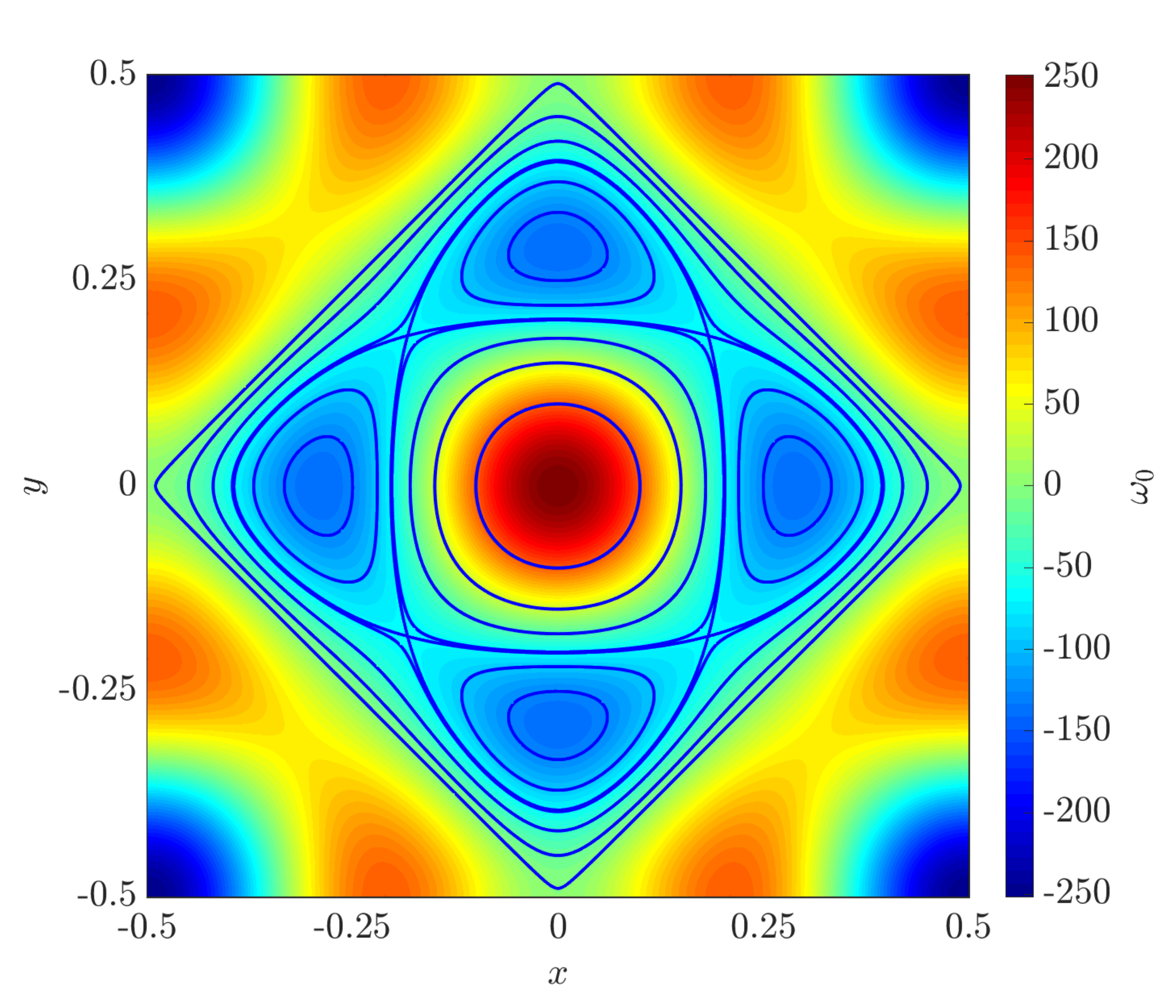}
		\par\end{centering}
	\caption{Streamlines of equation (\ref{eq:NS array}) with $\nu=0.001$ at
		$t=0$ overlaid on the initial vorticity field.\label{fig:vortex array}}
\end{figure}

The topology of the streamlines of the steady, inviscid velocity field $\mathbf{v}_{E}(\mathbf{x})$ is depicted in Fig. \ref{fig:vortex array}. The streamline geometry
of the unsteady solution $\mathbf{v}(\mathbf{x},t)$ remains the same. The central feature of this
flow is delineated by the heteroclinic connections between the hyperbolic
fixed points located at $\left(0,0.5\right),\left(0.5,0\right),\left(0,-0.5\right),\left(-0.5,0\right)$. 
This heteroclinic network encompasses an array of vortical recirculation regions around the elliptic fixed points located at $(0,0)$ and $\left(e,0\right),\left(-e,0\right),\left(0,e\right),\left(0,-e\right)$
with $e=\frac{1}{\pi}\arccos\left(\frac{\sqrt{6}}{4}\right)$. Furthermore,
these vortical domains are separated from each other and from the
outer heteroclinic network by an inner collection of heteroclinic
connections among the hyperbolic fixed points located at $\left(h,h\right),\left(-h,h\right),\left(h,-h\right),\left(-h,-h\right)$
with $h=\frac{1}{\pi}\arccos(\sqrt{\sqrt{6}/3+2}/2)$.

All closed, periodic streamlines in the vortical regions
are perceived as structures hindering the spread of high absolute
vorticity from the centers of the vortical regions. Moreover, the
periodic streamlines between the inner and the outer heteroclinic
network should also be deemed as barriers to the transport of vorticity.

\begin{figure}[H]
	\begin{centering}
		\includegraphics[width=1\textwidth]{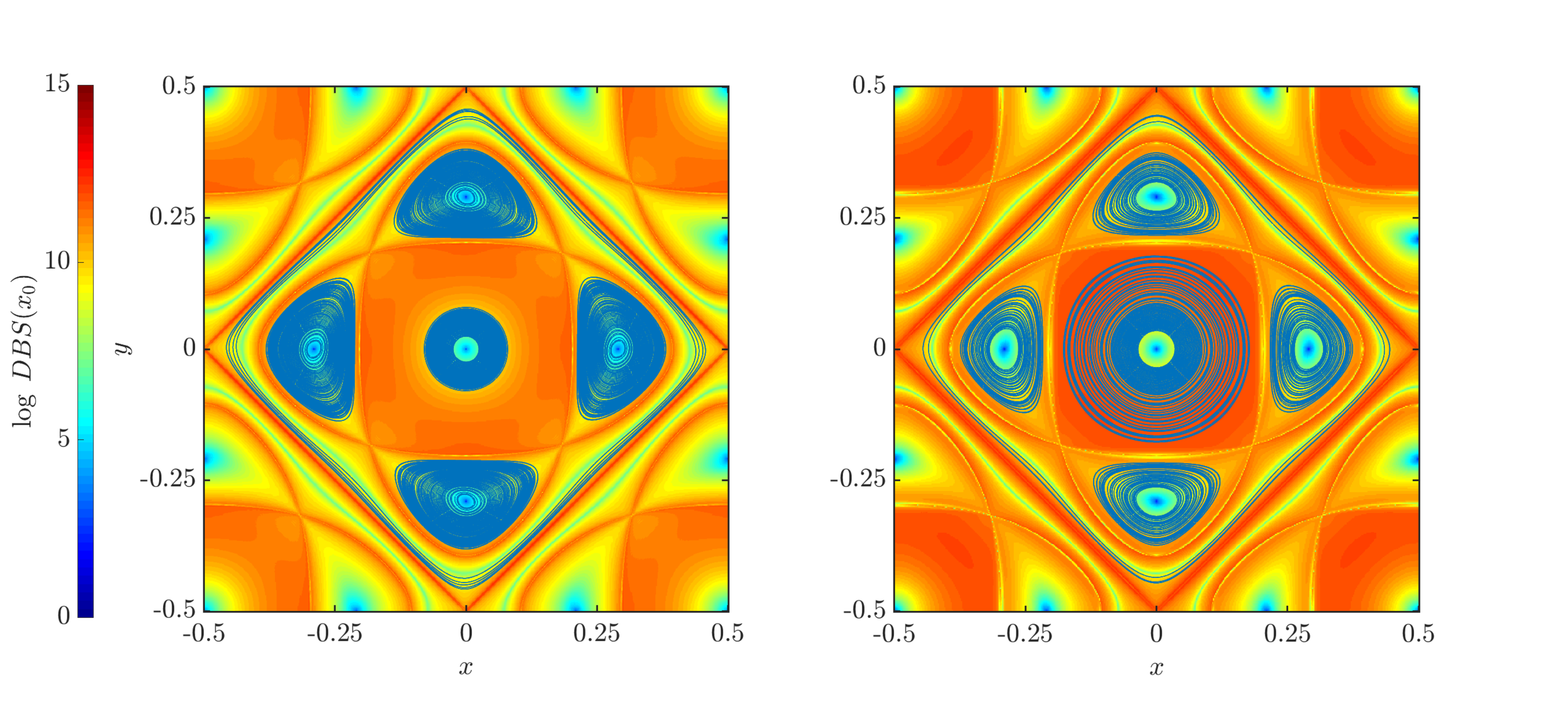}
		\par\end{centering}
	\caption{Barriers to vorticity transport (blue) superimposed on the $\text{DBS}_{t_{0}}^{t_{1}}(\mathbf{x}_{0})$
		field inside a vortex array of equation (\ref{eq:NS array}). The analysis was performed for $t_0=0$ and $t_{1}=1$ (left) and $t_1=5$ (right).
		In both cases, we set the kinematic viscosity $\nu=0.001$.
		\label{fig:vortex cross}}
\end{figure}

To verify this, we use Algorithm~\ref{alg:closedVortexBoundary}
to detect diffusive vortex boundaries based on Eq. (\ref{eq:NS array}) for
two different integration times. As we observe in fig. \ref{fig:vortex cross},
as the integration time $t_1-t_0$ increases,
the extracted diffusive vortex boundaries grow in number and size in the central
elliptic region of high vorticity, whereas they become tighter around
the four cores of high negative vorticity. 
Moreover, our algorithm captures larger diffusive vortex
boundaries that closely align with both the inner and outer heteroclinic
networks. Finally, we note the high correlation between the extracted diffusive
vortex boundaries and the ridges of the DBS field.

\subsection{Two-dimensional turbulence}
We use a standard pseudo-spectral code to solve the two-dimensional,
incompressible Navier-Stokes equations,

\[
\partial_{t}\mathbf{v}+\mathbf{v}\cdot\bm{\nabla}\mathbf{v}=-\bm{\nabla}p+\nu\Delta\mathbf{v},\qquad\bm{\nabla}\cdot\mathbf{v}=0,
\]
The domain is $[0,2\pi]\times[0,2\pi]$ with periodic boundary conditions.
At Reynolds number $Re=\nu^{-1}=5\times10^{4}$, the spatial coordinates
are resolved using $1024^{2}$ Fourier modes with $2/3$ dealiasing.
To construct the transport vector field, we advect trajectories from
an initial grid of $1024\times1024$ points over the time interval
$[0,50]$. The Runge--Kutta algorithm of MATLAB (i.e., ode45) is
used for the numerical integration. This algorithm uses an adaptive
time stepping such that the relative and absolute errors are below
$10^{-6}$.

\begin{figure}
	\begin{centering}
		\includegraphics[width=0.9\textwidth]{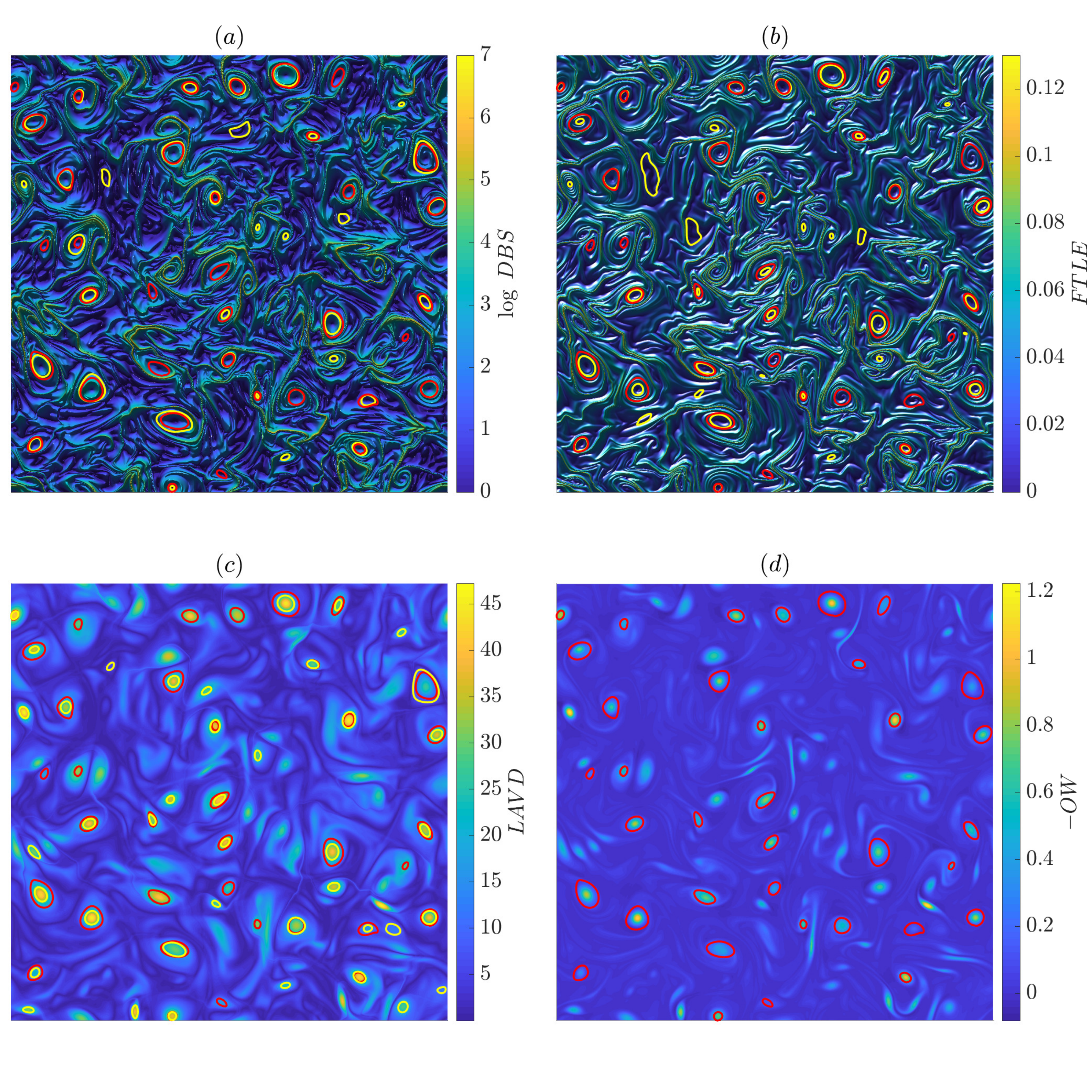}
		\par\end{centering}
	\caption{Lagrangian and Eulerian vortex identification methods on a decaying
		turbulence simulation with integration time $t_{1}=50$. (a) Diffusive vortex
		boundaries (red) and material diffusion barriers (yellow) superimposed
		with the $\mathrm{DBS}_{0}^{50}(\mathbf{x}_{0})$ field. (b) Diffusive vortex boundaries (red) and black-hole eddies (yellow) overlaid on the $FTLE(\mathbf{x}_{0})$
		field. (c) Diffusive vortex boundaries (red) and LAVD vortices (yellow) superimposed
		with the $LAVD(\mathbf{x}_{0})$ field. (d) Diffusive vortex boundaries (red)
		overlaid on the negative Okubo-Weiss parameter field. All panels show the entire computational domain $[0,2\pi]\times [0,2\pi]$. \label{fig:structures}
	}
\end{figure}

Fig. \ref{fig:structures} shows different Lagrangian and
Eulerian vortex identification methods for this computational experiment.
In the entirety of fig. \ref{fig:structures}, we denote with red
color the extracted diffusive vortex boundaries. For reference, we have also
used BarrierTool to compute outermost material barriers to passive
scalar diffusion \cite{haller18-1}, which are shown in yellow in
fig. \ref{fig:structures}(a) as well as black-hole eddies \cite{haller2013coherent},
which are shown in yellow in fig. \ref{fig:structures}(b). Moreover,
in fig. \ref{fig:structures}(c) we have identified Lagrangian-averaged
vorticity deviation (LAVD) eddies using the parameter values described
on \cite{haller16}. For each of these plots the overlaid scalar fields
correspond to the diagnostic field tied to the type of barriers that
we have extracted (DBS, FTLE and LAVD, respectively). For all the
Lagrangian vortex identification methods of fig. \ref{fig:structures}
we observe a clear correlation between all types of barriers, yet
there are regions that only admit one kind of barrier but none of
the others.

In addition, fig. \ref{fig:structures}(d) shows the negative
Okubo-Weiss (OW) parameter field. The OW parameter is defined as

\begin{equation}
\mathrm{OW}(\mathbf{x},t)=s_{2}^{2}(\mathbf{x},t)-\omega^{2}(\mathbf{x},t)
\end{equation}
where $s_{2}$ is the largest eigenvalue of the symmetric
part of the velocity gradient. This quantity is broadly used in the
literature to locate instantaneous vortical regions at domains where
it attains negative values \cite{okubo1970,weiss1991}.

\begin{figure}
	\begin{centering}
		\includegraphics[width=0.8\textwidth]{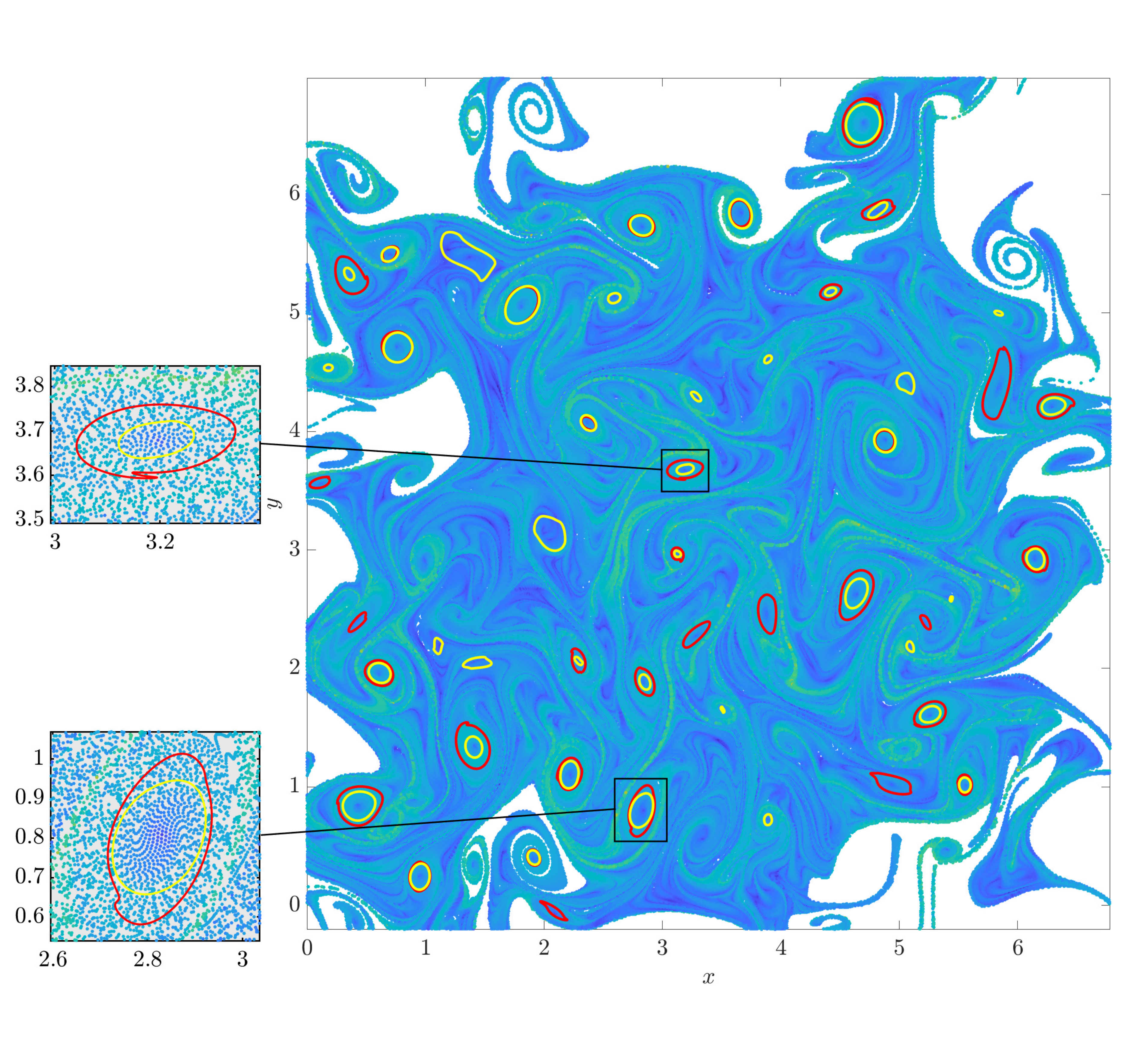}
		\par\end{centering}
	\caption{Final position at time $t_{1}=50$ of advected diffusive vortex boundaries (red)
		and black-hole eddies (yellow) overlaid on the advected position of
		a uniform grid of $500\times500$ tracers color-coded with their $\mathrm{DBS}_{0}^{50}(\mathbf{x}_{0})$
		value. Close-ups: Tangential filamentation of the diffusive vortex boundaries in contrast to the unstretched black-hole eddies. \label{fig:structuresAdvected}}
\end{figure}

Fig. \ref{fig:structuresAdvected} shows the final positions
of the extracted diffusive vortex boundaries (red) and black-hole eddies (yellow)
against the positions of an initially uniform grid of points color-coded
with their DBS value. Black-hole eddies show no filamentation in agreement
with their construction as locally minimally stretching coherent structures
\cite{haller2013coherent}. In constrast, diffusive vortex boundaries, constructed
as extremizers to the transport of vorticity, manifest tangential
stretching in some cases (blown up figures in Fig. \ref{fig:structuresAdvected}). However, transport is still efficiently
hindered by closed material curves that filament in tangential directions
without a global breakaway that creates smaller scales. Moreover,
the apparent dissimilarity in the detection of barriers in some regions
is explained by the initial vorticity distribution (a constraint in
our calculus of variations problem) which may tip the scales in favor
of or against the detection of diffusive vortex boundaries irrespective of the
existence of black-hole eddies.

\begin{figure}
	\begin{centering}
		\includegraphics[width=0.9\textwidth]{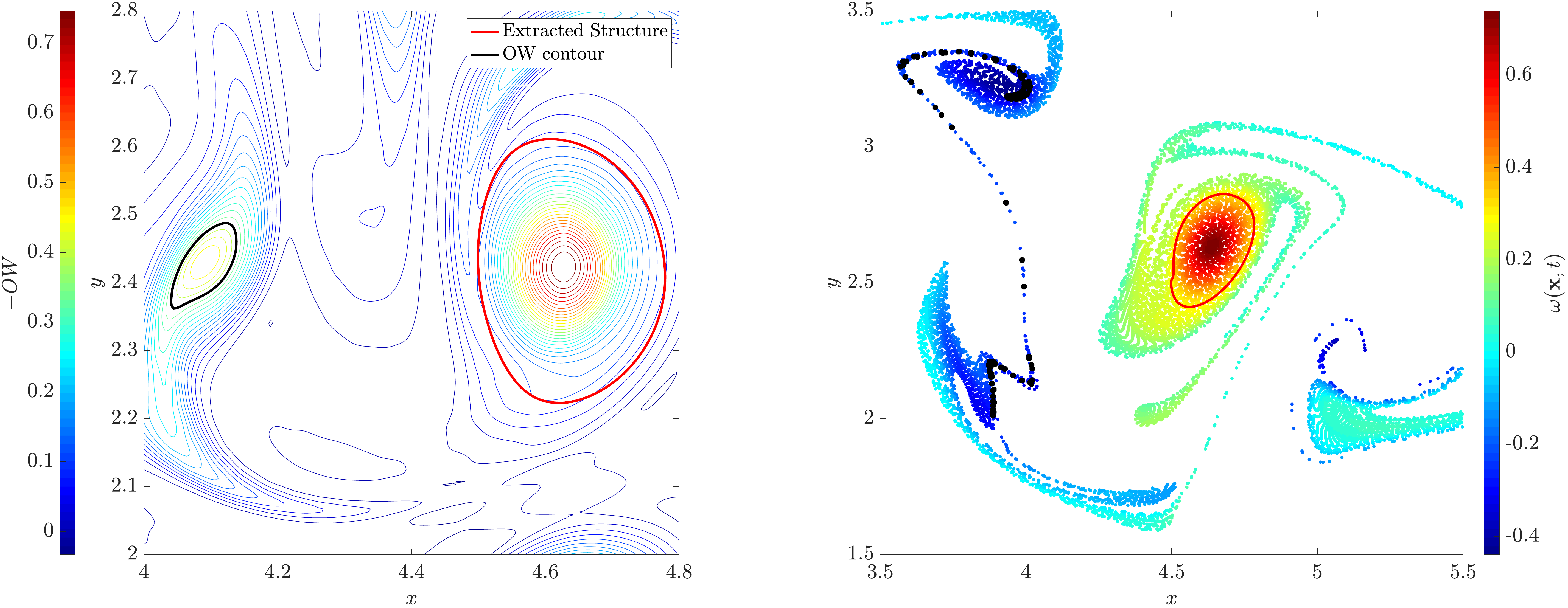}
		\par\end{centering}
	\caption{Left: Extracted diffusive vortex boundary (red) overlaid on level sets of the
		Okubo-Weiss (OW) parameter. Right: Advected image of the diffusive vortex boundary
		and a set of tracers lying initially on the black level set of the
		OW parameter superimposed with the final position of an originally
		uniform grid of points color-coded with their vorticity value. \label{fig:okubo}}
\end{figure}

The level curves of the OW parameter are often viewed as
coherent structures in the flow \cite{dubief2000coherent}. We investigate
this claim in fig. \ref{fig:okubo} which depicts a zoomed-in region
close to the center of the computational domain along with an extracted diffusive
vortex boundary, overlaid on level sets of the Okubo-Weiss (OW) parameter.
In this region, $\mathrm{OW}(\mathbf{x})$ signals two different vortical
regions, while our algorithm only locates a material vortex boundary
(as outermost barrier to vorticity transport) in one of these regions.
To examine this prediction more closely, we compare the advected image
of a set of tracers seeded along a level set of the $\mathrm{OW}(\mathbf{x},t)$
parameter against the final position of the diffusive vortex boundary. We observe
that the material region obtained from our algorithm remains a coherent
vortex that keeps vorticity concentrated. Over the same time interval,
the material region surrounded by the OW level set completely falls
apart, and hence this level set fails to prevent vorticity from leaking
out from a coherent core.

\begin{figure}
	\begin{centering}
		\includegraphics[width=1\textwidth]{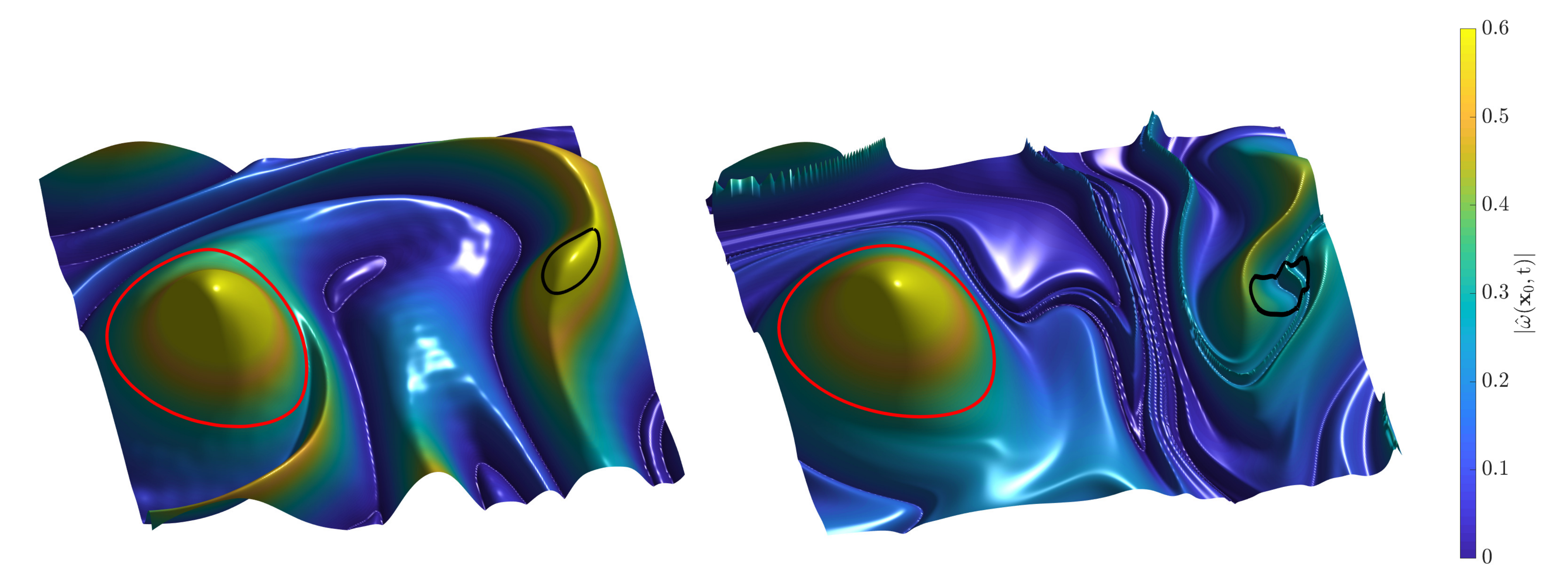}
		\par\end{centering}
	\caption{Evolution of $|\hat{\omega}(\mathbf{x}_{0},t)|$, the vorticity norm
		in Lagrangian coordinates. Left: Colored contours of $|\hat{\omega}(\mathbf{x}_{0},0)|$
		with the diffusive vortex boundary extracted from our algorithm (red) and OW
		level set (black). Right: Contours of $|\hat{\omega}(\mathbf{x}_{0},50)|$
		with the same red and black curves at $t_{1}=50$. \label{fig:3D}
	}
\end{figure}

Finally, the same image depicted in Lagrangian coordinates
is shown in Fig. \ref{fig:3D}. More specifically, the norm of vorticity
is portrayed as a surface over the Lagrangian coordinates $\mathbf{x}_{0}$
for two different configurations ($t_{0}=0$ and $t_{1}=50$). We
observe that along the extracted diffusive vortex boundary vorticity is diffused
in a uniform fashion which is in agreement with the underlying variational
principle (minimal vorticity leakage) used in its construction. In
contrast, the OW level set indicates no such organizing role in the
vorticity landscape showing preferential directions along which vorticity
diffuses more compared to the rest of the curve.

\section{Conclusions}
We have proposed defining two-dimensional vortices as maximal
regions enclosed by material barriers to the viscous transport of
vorticity in Navier--Stokes flows. With this approach, we have been
able to leverage the two-dimensional version of recent results of Haller et al.
\cite{haller18-2} on strongest material barriers to diffusion of
a general passive scalar. We have used the conservation law provided
by that theory to derive a three-dimensional, autonomous system of
ODEs. Outermost closed projections of the orbits of this ODE onto
the space of Lagrangian positions mark material curves satisfying
our diffusive vortex boundary definition. Although not a part of the current
work, a similar definition involving open solutions to the constrained
barrier equations can be used to reveal signatures of material jet
cores and fronts in the vorticity field. An extension of the present
results to three dimensions, however, will require major modifications
since three-dimensional vorticity is a vectorial quantity and no longer satisfies a linear advection-diffusion equation for a known velocity field.

We have also introduced a numerical algorithm that automatizes
the proposed vortex identification procedure. Upon comparing our algorithm
with different Lagrangian vortex detection methods (geodesic theory
of LCSs, LAVD, PRA) we find the present algorithm to be more computationally
expensive owing to the need for computing the flow map gradient and
the spatial derivatives of vorticity. At the same time, the proposed
algorithm is objective (observer-independent), takes the Navier--Stokes
vorticity dynamics into account, and requires no reliance on user
input or heuristic parameters. Our open source MATLAB package, BarrierTool,
provides a full implementation of the present results, as well as
implementations of other Lagrangian vortex detection methods that
are based on various material coherence principles.

\begin{acknowledgments}
We are grateful to Daniel Karrasch and Hessam Babaee for useful
discussions and comments. S.K. and G.H. acknowledge support from the
Turbulent Superstructures priority program of the German National
Science Foundation (DFG).
\end{acknowledgments}

\bibliography{DiffusiveVortexBoundaries}

\end{document}